\documentclass[%
 reprint,
 superscriptaddress,preprintnumbers,
 nofootinbib,
 amssymb,
 aps,
]{revtex4-1}

\usepackage[utf8]{inputenc}
\usepackage[colorlinks=true,citecolor=blue,linkcolor=blue]{hyperref}
\usepackage[normalem]{ulem}
\usepackage{amsmath,amssymb,mathtools}
\usepackage{epsfig}
\usepackage{graphicx}  
\usepackage{booktabs}
\usepackage{url}
\usepackage{color}
\usepackage{slashed}
\usepackage{cancel}
\usepackage{multirow}
\usepackage{placeins}
\usepackage[dvipsnames]{xcolor}
\usepackage{epstopdf}
\usepackage{soul}
\usepackage{tikz}
\usepackage[capitalise, english]{cleveref}
\usepackage{siunitx}
\usepackage{xspace}
\usepackage{lipsum}
\usetikzlibrary{trees}
\usetikzlibrary{decorations.pathmorphing}
\usetikzlibrary{decorations.markings}

\definecolor{red}{rgb}{0.6,.0706,.1373}
\definecolor{blue}{rgb}{0,0.396,0.741}
\newcommand\myshade{80}
\colorlet{mylinkcolor}{violet}
\colorlet{mycitecolor}{violet}
\colorlet{myurlcolor}{violet}

\newcommand{\mU}{m_U^2}

\hypersetup{
  linkcolor  = mylinkcolor!\myshade!black,
  citecolor  = mycitecolor!\myshade!black,
  urlcolor   = myurlcolor!\myshade!black,
  colorlinks = true
}

\allowdisplaybreaks

\newcommand{\be}{\begin{equation}}
\newcommand{\ee}{\end{equation}}
\newcommand{\bea}{\begin{eqnarray}}
\newcommand{\eea}{\end{eqnarray}}

\def\beq#1\eeq{\begin{align}#1\end{align}}

\makeatletter
\providecommand*{\diff}%
  {\@ifnextchar^{\DIfF}{\DIfF^{}}}
\def\DIfF^#1{%
  \mathop{\mathrm{\mathstrut d}}%
    \nolimits^{#1}\gobblespace}
\def\gobblespace{%
  \futurelet\diffarg\opspace}
\def\opspace{%
  \let\DiffSpace\!%
  \ifx\diffarg(%
    \let\DiffSpace\relax
  \else
    \ifx\diffarg[%
      \let\DiffSpace\relax
    \else
        \ifx\diffarg\{%
        \let\DiffSpace\relax
      \fi\fi\fi\DiffSpace}

\keywords{}

\begin{document}

\title{Third-Family Lepton-Quark \emph{Fusion}}

\author{Arman Korajac}
\email{arman.korajac@ijs.si}
\affiliation{J. Stefan Institute, Jamova 39, P. O. Box 3000, 1001 Ljubljana, Slovenia}
\author{Peter Krack}
\email{peter.krack@nikhef.nl}
\affiliation{Department of Physics and Astronomy, Vrije Universiteit, NL-1081 HV Amsterdam}
\affiliation{Nikhef Theory Group, Science Park 105, 1098 XG Amsterdam, The Netherlands}
\author{Nud\v{z}eim Selimovi\'c}
\email{nudzeim.selimovic@pd.infn.it}
\affiliation{Istituto Nazionale di Fisica Nucleare, Sezione di Padova, Via Francesco Marzolo 8, 35131 PD, Italy}

\preprint{Nikhef 2023-019}

\begin{abstract}
We analyze the signatures of new physics scenarios featuring third-family quark-lepton unification at the TeV scale in lepton-quark fusion at hadron colliders. Working with complete UV dynamics based on the $SU(4)$ gauge symmetry in the third-family fermions, we simulate the resonant production of a vector leptoquark at the next-to-leading order, including its decay and matching to the parton showers. The precise theoretical control over this production channel allows us to set robust bounds on the vector leptoquark parameter space which are complementary to the other production channels at colliders. We emphasize the importance of the resonant channel in future searches and discuss the impact of variations in the model space depending on the flavor structure of the vector leptoquark couplings.
\end{abstract}

\maketitle

\section{Introduction} \label{sec:intro}
The experiments probing the Standard Model (SM) at different energies impose considerable constraints on new physics (NP) models addressing the Higgs hierarchy problem at the TeV scale. Notably, the present high-energy bounds on NP show significant variations depending on the flavor structure of possible new states. In particular, if the new states couple universally to the SM fermions with $\mathcal{O}(1)$ couplings, searches at the Large Hadron Collider (LHC) push their mass in the multi-TeV range. On the contrary, if the new states are primarily coupled to the third family, the LHC bounds on their mass become prominently weaker, often not surpassing $1$ TeV~\cite{Allwicher:2023shc}. The flavor structure of such NP, mainly coupled to the third family, evokes the connection to the hierarchical SM Yukawa couplings that, to a first approximation, can be described by nonvanishing entries corresponding to the interactions of the Higgs with the third-family fermions. Consequently, such a flavor non-universal NP could be a first step in addressing the SM flavor puzzle. Furthermore, linking NP at the TeV scale to the solution of the Higgs hierarchy problem opens as an enticing possibility.

The recent study in~\cite{Davighi:2023iks} provided a comprehensive investigation of NP models to elucidate the emergence of flavor hierarchies within a flavor non-universal gauge framework. It was found that, in order to mitigate significant quantum corrections to the Higgs mass, being consistent with the ``finite naturalness'' criterion~\cite{Farina:2013mla}, it is crucial for the initial layer of non-universality distinguishing the third family from the lighter families, to manifest itself at the TeV scale. Moreover, adding further criteria encompassing consistency with experimental data, and semi-simple embedding in the ultraviolet, filtered out a few models as possible SM extensions. Remarkably, all viable models exhibit quark-lepton unification within the third family together with a non-universal electroweak sector. 

The possible intermediate step, before reaching the SM, common to all the phenomenologically viable models featuring the third-family quark-lepton unification is the so-called 4321 gauge group, $\mathcal{G}_{4321}=SU(4)_{3}\times SU(3)_{12}\times SU(2)_L\times U(1)_X$, at the TeV scale~\cite{DiLuzio:2017vat,Greljo:2018tuh}. The subscripts in $SU(4)_{3}\times SU(3)_{12}$ denote the flavor non-universal scenario and indicate that the would-be third-family fermions are charged under $SU(4)_3$, while light families are charged under $SU(3)_{12}$. The gauge group $SU(2)_L$ is flavor universal and acts as in the SM, while $U(1)_X$ coincides with the SM hypercharge for light families and $T_R^3$ for the third family. 

Aside from its theoretical appeal in providing a way to address the SM flavor puzzle and the Higgs hierarchy problem~\cite{Bordone:2017bld,Fuentes-Martin:2020bnh,Fuentes-Martin:2020pww,Fuentes-Martin:2022xnb}, the quark-lepton unification of this type necessarily involves a TeV-scale $U_1$ vector leptoquark mainly coupled to the third family. Indeed, such a field has been recognized as a solution for the experimental hints of Lepton Flavor Universality (LFU) violation in
semileptonic B-meson decays with a charged current ($b\to c\tau\nu$) underlying quark transition~\cite{Buttazzo:2017ixm,BaBar:2012obs,BaBar:2013mob,Belle:2015qfa,Belle:2017ilt,Belle:2019rba,LHCb:2015gmp,LHCb:2023abc}. Moreover, it could explain tensions in B-meson branching ratios and differential distributions~\cite{LHCb:2014cxe,LHCb:2016ykl,LHCb:2015wdu,LHCb:2015tgy,LHCb:2020lmf,LHCb:2020gog,ATLAS:2018gqc,CMS:2017ivg,Belle:2016fev} through a non-standard LFU contribution to a neutral current ($b\to s\ell\ell$) amplitude~\cite{Bobeth:2011st,Crivellin:2018yvo,Angelescu:2018tyl, Isidori:2023unk}. In this context, thorough studies of the implications of addressing B-anomalies via $U_1$ for $\Delta F=1$ (including the impact in $B\to K\nu\bar{\nu}$ transitions, interesting in light of the recent Belle II measurement~\cite{belleIItalk2023}) and $\Delta F=2$ transitions~\cite{DiLuzio:2018zxy,Fuentes-Martin:2020hvc,Crosas:2022quq}, $\tau$ decays~\cite{Allwicher:2021ndi}, and electroweak precision observables (EWPO)~\cite{Allwicher:2023aql}, revealed consistency with $\mathcal{G}_{4321}$ being at the TeV scale. 

Regarding the direct probes of leptoquark states at the LHC, 
experimental collaborations rely on three well-established processes. First, there is a leptoquark pair production in a $gg$ or a $q\overline{q}$ fusion via strong interactions, or a lepton exchange in $t$-channel~\cite{Bessaa:2014jya,Dorsner:2014axa}. Being strongly phase-space suppressed, a leptoquark pair production is not optimal for heavy leptoquark searches. Second, there is a single leptoquark plus lepton production in a quark-gluon scattering~\cite{Alves:2002tj,Hammett:2015sea,Mandal:2015vfa,Dorsner:2018ynv}. It suffers less phase-space suppression, and for
$\mathcal{O}(1)$ couplings, it surpasses pair
production. Third, a non-resonant $t$-channel leptoquark exchange in a Drell-Yan process
could manifest as a deviation in the high-$p_T$ tail of the dilepton invariant mass distribution~\cite{Faroughy:2016osc, Schmaltz:2018nls, Greljo:2017vvb,Fuentes-Martin:2020lea, Greljo:2018tzh,Marzocca:2020ueu, Baker:2019sli, Haisch:2022afh, Haisch:2022lkt}. In the limit of large couplings and masses beyond the kinematical reach for on-shell production, this process outweighs the other two. For recent experimental results, we refer the reader to~\cite{CMS:2020wzx, ATLAS:2022fho,ATLAS:2023vxj,CMS:2023qdw}.

In addition to the aforementioned search strategies, there has been a recent interest in the resonant leptoquark production from a direct lepton-quark fusion. The initial idea was put forth in~\cite{Ohnemus:1994xf}, while practical implementations became possible only after the precise determination of
lepton parton distribution functions (PDFs) inside the proton~\cite{Buonocore:2020nai, Manohar:2016nzj, Manohar:2017eqh, Almeida:2022udp, Ajmal:2023yhl}. The suppression from lepton PDFs is compensated by the least
phase-space suppression, making this channel competitive with other production mechanisms.
Indeed, the previous work on this topic concerning scalar leptoquarks~\cite{Buonocore:2022msy} showed that the lepton-quark fusion channel should be utilized as a complementary probe in parameter regions with $\mathcal{O}(1)$ leptoquark couplings and TeV-scale masses. Interestingly, this parameter region coincides with the one for the $U_1$ leptoquark in 4321 constructions, calling for a detailed analysis of this production channel in a complete UV model.
\\\indent In fact, an analysis has been performed at leading order (LO) in the NP coupling, where the authors examined the resonant $U_1$ production at the LHC and its exclusion power on the $U_1$ parameter space~\cite{Haisch:2020xjd}. At the time of conducting this analysis, no experimental searches for the resonant leptoquark production were available, hence the limits were derived using Monte Carlo-generated samples. Additionally, the signal events were showered by \texttt{Pythia}~\cite{Sjostrand:2014zea} which at present does not handle initial-state radiation from leptons. To circumvent this issue, initial leptons were substituted for photons in the \texttt{MadGraph}~\cite{Alwall:2014hca} generated LHE files \cite{Buonocore:2020erb}. 

We expand on this study in three ways. First, we calculate the NLO QCD + QED corrections to the $U_1$ production cross-section. Second, we match to \texttt{Herwig} which has an implemented lepton-showering algorithm~\cite{Bellm:2019zci} via the $\texttt{POWHEG}$ method~\cite{Alioli:2010xd}. A study involving scalar leptoquarks ~\cite{Buonocore:2022msy} found that a proper treatment of initial-state radiation by leptons introduces an approximate 15\% difference with respect to the particle label manipulation as is the case with \texttt{Pythia} in~\cite{Haisch:2020xjd}. Finally, the CMS collaboration recently reported the first search for leptoquarks produced in a lepton-quark collision~\cite{CMS:2023bdh}, which we utilize to find exclusion limits on the $U_1$ leptoquark coupling for different values of its mass. 

The paper is structured as follows: In Sec.~\ref{sec:model}, we explain the 4321 model and the $\mathcal{G}_{4321}\to \rm{SM}$ breaking pattern. We derive the relevant $U_1$ leptoquark Lagrangian and discuss its flavor structure. In Sec.~\ref{sec:resProd}, we present the details of the NLO calculation. Finally, Sec.~\ref{sec:pheno} is devoted to a phenomenological study from which we derive the bounds on the $U_1$ parameters using the lepton-quark fusion channel.

\section{The model}\label{sec:model}
\begin{table}[t]
\begin{center}
\setlength{\tabcolsep}{5pt}
\renewcommand{\arraystretch}{1.1}
\begin{tabular}{|c|c|c|c|c|c|c|}
\hline
Field & $SU(4)_3$ & $SU(3)_{12}$ &  $SU(2)_L$  & $U(1)_X$ \\
\hline
\hline 
 $\psi_L$ & $\mathbf{4}$  & $\mathbf{1}$  & $\mathbf{2}$  &  0 \\
$\psi_R^+$ & $\mathbf{4}$ &  $\mathbf{1}$   & $\mathbf{1}$  &  $ 1/2$  \\[3pt] 
$\psi_R^-$ & $\mathbf{4}$ &  $\mathbf{1}$   & $\mathbf{1}$  &  $-1/2$  \\[3pt] 
$q_L^i$ & $\mathbf{1}$ &  $\mathbf{3}$   & $\mathbf{2}$  &  $1/6$   \\[3pt] 
$\ell_L^i$ & $\mathbf{1}$ &  $\mathbf{1}$   & $\mathbf{2}$  &  $-1/2$  \\[3pt]
$u_R^i$ & $\mathbf{1}$ &  $\mathbf{3}$   & $\mathbf{1}$  &  $2/3$  \\[3pt] 
$d_R^i$ & $\mathbf{1}$ &  $\mathbf{3}$   & $\mathbf{1}$  &  $-1/3$  \\[3pt]  
$e_R^i$ & $\mathbf{1}$ &  $\mathbf{1}$   & $\mathbf{1}$  &  $-1$  \\[3pt]  
\hline
\hline
$H$ & $\mathbf{1}$ &  $\mathbf{1}$   & $\mathbf{2}$  &  $1/2$  \\
$\Omega_3$ & $\mathbf{\bar 4}$ &  $\mathbf{3}$   & $\mathbf{1}$  &  $1/6$  \\
$\Omega_1$ & $\mathbf{\bar 4}$ &  $\mathbf{1}$   & $\mathbf{1}$  &  $-1/2$  \\  
\hline
\end{tabular}
\end{center}
\caption{Field content and charges under the $\mathcal{G}_{4321}=SU(4)_3 \times SU(3)_{12} \times SU(2)_L\times U(1)_{X}$ gauge group. The index $i=1,2$ labels the light fermion families, and $\psi_L\equiv(q_L^3\; \ell_L^3)^\intercal$, $\psi_R^+\equiv(u_R^3\; \nu_R^ 3)^\intercal$ and $\psi_R^-\equiv(d_R^3\; e_R^3)^\intercal$ are multiplets unifying quarks and leptons of the third family.}
\label{tab:fields}
\end{table}

The model we consider is based on the $\mathcal{G}_{4321}$ gauge group and the matter content with transformation properties shown in Tab.~\ref{tab:fields}. The gauge bosons associated with $SU(4)_{3}\times SU(3)_{12}\times SU(2)_L\times U(1)_X$ dynamics are denoted by $H_\mu^{A}$, $C_\mu^a$, $W_\mu^I$, and $X_\mu$, respectively. The corresponding adjoint indices are $A=1,\dots,15$, $a=1,\dots,8$, and $I=1,2,3$, while the corresponding gauge couplings are $g_4$, $g_3$, $g_2$, and $g_1$.

The spontaneous symmetry breaking of $\mathcal{G}_{4321}$ to $\mathcal{G}_{\rm SM}=SU(3)\times SU(2)_L\times U(1)_Y$ results from the vacuum expectation values of two scalars $\Omega_3$ and $\Omega_1$ whose transformation properties under $\mathcal{G}_{4321}$ are shown in Tab.~\ref{tab:fields}. The vacuum direction is such that $\mathcal{G}_{\rm SM}\supset SU(3)\times U(1)_Y=\left[SU(4)_{3}\times SU(3)_{12}\times U(1)_X\right]_{\rm diag}$, and the hypercharge is given by $Y=X+\sqrt{2/3}\,T_4^{15}$, where $T_4^{15}=\frac{1}{2\sqrt{6}}{\rm diag}(1,1,1,-3)$ is the $SU(4)_3$ generator. The massive gauge bosons resulting from the $\mathcal{G}_{4321}\to\mathcal{G}_{\rm SM}$ breaking include the coloron $G'$, the $U_1$ vector leptoquark, and the neutral $Z'$, which transform under $\mathcal{G}_{\rm SM}$ as $G'\sim({\bf 8},{\bf 1},0)$, $U_1\sim({\bf 3},{\bf 1},2/3)$, and $Z'\sim({\bf 1},{\bf 1},0)$. They correspond to the following linear combinations of the $\mathcal{G}_{4321}$ gauge eigenstates
\begin{equation}
     \label{eq:masseigen}
     \begin{aligned}   
    G_\mu^{\prime a} &= c_3 H_\mu^a - s_3 C_\mu^a\,,\quad Z_\mu^{\prime} = c_1 H_\mu^{15} - s_1 X_\mu\,,\\
    U_\mu^{1,2,3} &= \frac{1}{\sqrt{2}}(H_\mu^{9,11,13}-iH_\mu^{10,12,14})\,,
\end{aligned}
\end{equation}
where the mixing angles are $\theta_1 = \arctan(\sqrt{2/3}\,g_1/g_4$), $\theta_3 = \arctan(g_3/g_4)$, and we used $c_{1,3} \equiv \cos\theta_{1,3}$ and $s_{1,3} \equiv \sin\theta_{1,3}$. The masses they obtain read
\begin{equation}
    m_{U} = \frac{1}{2}g_4 f_U\,, \quad m_{Z',G'} = \frac{m_U}{c_{1,3}} \frac{f_{Z',G'}}{f_U}\,,
    \label{eq:masses}
\end{equation}
where $f_U^2 = \omega_1^2+\omega_3^2$, $f_{Z'}^2 = 3\omega_1^2/2+\omega_3^2/2$, $f_{G'}^2 = 2\omega_3^2$, and $\omega_{1,3}$ denote the vacuum expectation values of $\Omega_{1,3}$. The field combinations orthogonal to $G^\prime$ and $Z^\prime$  in Eq.~\eqref{eq:masseigen} correspond to the SM gluons, $G^a_\mu$, and the hypercharge gauge boson, $B_\mu$,
\begin{equation}
\begin{aligned}
G^a_\mu&= ( s_3\, H^a_\mu + c_3\,C^a_\mu)\,,\phantom{aaa} 
g_s = g_3\, c_3 = g_4\, s_3\,,\\
B_\mu&= ( s_1\, H^{15}_\mu + c_1\,X_\mu)\,,\phantom{aa}
g_Y = g_X\, c_1 = \sqrt{\frac{3}{2}}g_4\, s_1\,,
\end{aligned}
\end{equation}
where we also expressed the QCD coupling, $g_s$, and the hypercharge coupling $g_Y$ in terms of $\mathcal{G}_{4321}$ gauge couplings and mixing angles.

In this paper, we are interested in the production of $U_1$ leptoquark by lepton-quark fusion at hadron colliders, taking into account the NLO corrections from its interactions with gluons and photons. Thus, the relevant interactions for our purpose are described by the following Lagrangian
\begin{equation}
    \begin{aligned}\label{eq:U1Lag}
    \mathcal{L}_{U_1} &\supset   \frac{g_4}{\sqrt{2}}\, U_\mu\left(\beta_L^{i\alpha}\,\overline{q}_L^i\gamma^\mu\ell_L^\alpha + \beta_R^{i\alpha}\,\overline{q}_R^i\gamma^\mu\ell_R^\alpha + {\rm h.c.}\right)\\
    & - i g_s U_\mu^\dagger T^a U_\nu G^{a,\mu\nu}-\frac{2}{3}ie \,U_{\mu}^\dagger U_{\nu} F^{\mu\nu}\,,
\end{aligned}
\end{equation}
where $G_{\mu\nu}^{a}$ and $F_{\mu\nu}$ are the gluon and the photon field strength tensors, and $e$ is the QED coupling. The couplings to the left- and right-handed fermions $\beta_L$ and $\beta_R$ are $3\times 3$ matrices in the quark-lepton family space that encode the flavor structure of the $U_1$ interactions. In the minimal version of the model, the $U_1$ couples to the third-family fermions only with $|\beta_{L,R}^{33}|=1$, resulting in the exact $U(2)^5$ flavor symmetry of its interactions. However, in more realistic models, one expects populating the other entries of $\beta_L$ following the $U(2)^5$ symmetry breaking~\cite{Barbieri:2015yvd,Fuentes-Martin:2019mun}. The low-energy phenomenology of the $U_1$ leptoquark constrains the size of this breaking, pointing to suppression in the $\beta_L$ entries parametrizing the couplings to light-family fields~\cite{Fuentes-Martin:2020hvc,Crosas:2022quq,Fuentes-Martin:2019mun}. Thus, a dominant resonant production of the $U_1$ proceeds through the $b+\tau$ collisions, as the PDF enhancement from lighter quarks is insufficient to compensate for their suppressed couplings to $U_1$. In Sec.~\ref{subsec:Projected_bounds}, we have explicitly checked that turning on couplings to light families compatible with all constraints leads to a negligible change in the exclusion bounds in the $U_1$ parameter space. Nevertheless, in the Monte Carlo tool we built, we keep $\beta_{L,R}$ as generic matrices and allow users to choose any values in order to match to other model implementations.
\section{\texorpdfstring{$U_1$}{U\_1} leptoquark resonant production @ NLO}\label{sec:resProd}
In order to characterize the $U_1$ vector leptoquark production in lepton-quark collisions as precisely as possible, we rely on the Monte Carlo tool developed in~\cite{Buonocore:2022msy}. In short, it is a \texttt{POWHEG-BOX-RES} implementation of the resonant leptoquark production apt of generating events which are processed by \texttt{Herwig} in order to obtain a full simulation of the collision at the NLO+PS level. We refer to Ref.~\cite{Buonocore:2022msy} for the details of the implementation, reminding that in general, it requires the computation of the averaged matrix-squared elements for all real-radiation contributions, and the finite part of the virtual corrections computed in $\overline{\mathrm{MS}}$ scheme. 

\begin{figure}[t]
    \centering
    \includegraphics[width=0.45\textwidth]{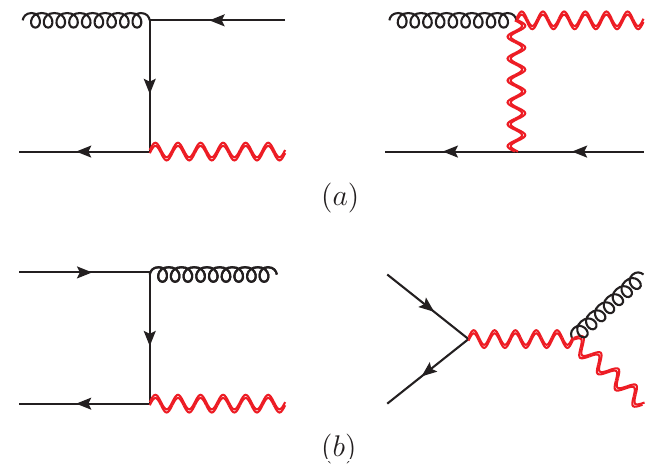}
    \caption{Real-radiation QCD corrections: a) gluon-initiated production $g + \ell\to q+ U_1$; b) gluon emission $q + \ell\to g+ U_1$. The $U_1$ vector leptoquark is shown in red.}
    \label{fig:QCD_real}
\end{figure}

In the following, we summarize the necessary ingredients specific for the case of $U_1$ vector leptoquark originating from $\mathcal{G}_{4321}$ symmetry breaking. First, the partonic Born cross-section for $q^i + \ell^\alpha \to U_1$ is given as
\begin{equation}\label{eq:bornU1}
    \hat{\sigma}_{\rm LO} = \frac{g_4^2}{4}(|\beta_L^{i\alpha}|^2+|\beta_R^{i\alpha}|^2)\,m_U^2\,.
\end{equation}

Second, the NLO QCD effects affecting the $U_1$ leptoquark production by lepton-quark fusion consist of real-radiation corrections involving:
the gluon-initiated production, $g(p_1) + \ell(p_2)\to q(k)+ U_1(q)$ in Fig.~\ref{fig:QCD_real}a), and the soft-gluon emission, $q(p_1) + \ell(p_2)\to g(k)+ U_1(q)$ in Fig.~\ref{fig:QCD_real}b), where the four-momenta of the particles are in the parentheses. The averaged matrix-squared element for the two processes read
\begin{align}
    \label{eq:g+lQCD} |\overline{\mathcal{M}}|^2_{g\ell} &= -\frac{1}{2} g_s^2 g_4^2\,\frac{s(s^2+t^2+2u \,\mU)}{t(s+t)^2}\,,\\
    \label{eq:q+lQCD}
    |\overline{\mathcal{M}}|^2_{q\ell} &= \frac{4}{3} g_s^2 g_4^2\,\frac{u(u^2+t^2+2s\mU)}{t(u+t)^2}\,,
\end{align}
where $s\,,t$, and $u$ are the partonic-level Mandelstam variables defined as $s=(p_1+p_2)^2$, $t=(p_1-k)^2$, $u= (p_1-q)^2$, and we kept only the $U_1$ mass $m_U$. 

\begin{figure}[t]
    \centering
    \includegraphics[width=0.48\textwidth]{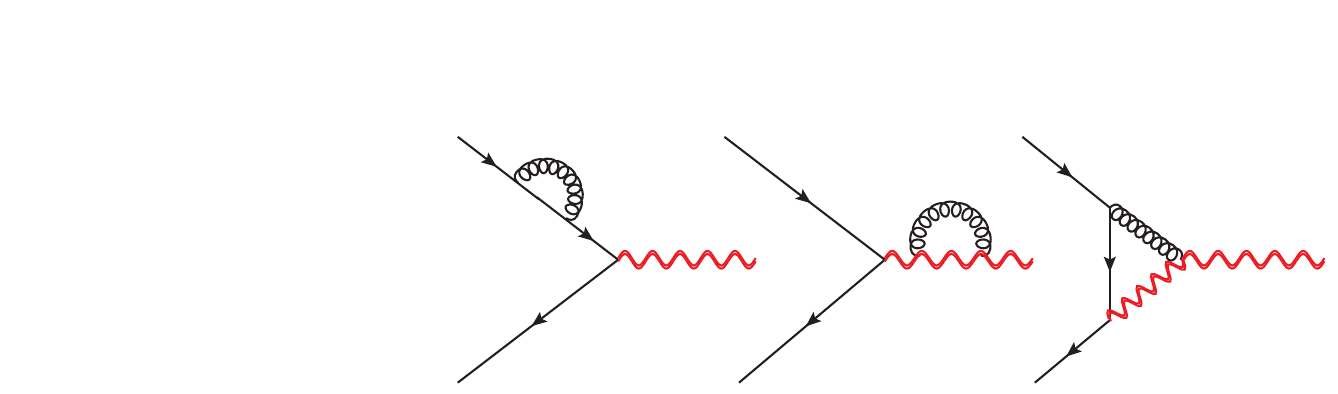}
    \caption{Virtual QCD corrections: quark and $U_1$ leg corrections, and the vertex correction. The $U_1$ vector leptoquark is shown in red.}
    \label{fig:QCD_virtual}
\end{figure}

Third, there are virtual QCD corrections originating from diagrams in Fig.~\ref{fig:QCD_virtual}. The details of evaluating the virtual NLO QCD corrections are presented in~\cite{Fuentes-Martin:2020luw}. The finite part of the virtual corrections computed in dimensional regularization that needs to be provided to \texttt{POWHEG}~\cite{Alioli:2010xd} reads
\begin{align}
    \label{eq:virtPWHG}
    \mathcal{V}_{\rm fin}&=  \frac{2}{3}g_4^2 s\left(\frac{13}{12}-\frac{4\pi}{\sqrt{3}}+\frac{\pi^2}{2}-\frac{5}{2}L_R-\frac{1}{2}L_R^2\right)\,,
\end{align}
where $L_R=\log(\mu_R^2/s)$, and $\mu_R$ is the renormalization scale. We note that consistently evaluating these corrections involves the effects of coloron states, $G^{\prime a}$, specific to the full model implementation. Here, we assumed that coloron and $U_1$ leptoquark are mass degenerate. We have also checked that taking the extreme case $m_{G'}=\sqrt{2}\,m_U$, see Eq.~\eqref{eq:masses}, has a negligible impact on our results.

\begin{figure}[t]
    \centering
    \includegraphics[width=0.46\textwidth]{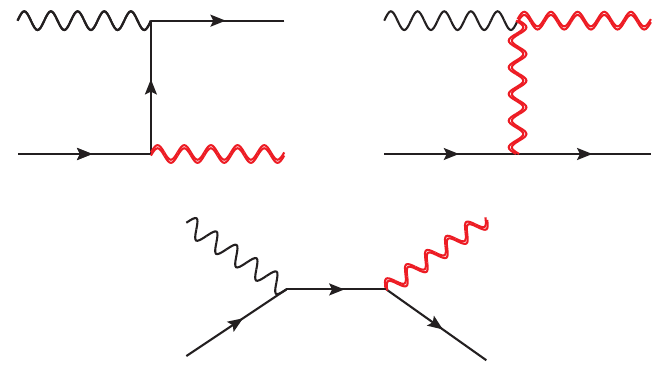}
    \caption{Real-radiation QED corrections $\gamma + q \to \ell + U_1$. The photon/lepton PDF enhancement compensates for the $\alpha_{QED}/\alpha_s$ suppression. The $U_1$ vector leptoquark is shown in red.}
    \label{fig:QED_real}
\end{figure}

In addition to the corrections that originate from the $U_1$ interactions with gluons, there are QED corrections to the resonant leptoquark production from the process initiated by the photon. 
Specifically, $\gamma(p_1)+q(p_2)\to \ell(k)+U_1(q)$ in Fig.~\ref{fig:QED_real}, contributes at the same order as the QCD corrections due to photon to lepton PDF enhancement compensating for the $\alpha_s$ to $\alpha_{\rm QED}$ suppression. The averaged matrix-squared element for this process reads

\begin{equation}
    |\overline{\mathcal{M}}|^2_{\gamma q}\, = -e^2g_4^2 \bigg(\frac{Q_{\ell}s+Q_{\rm q}t}{s+t}\bigg)^2 \frac{s^2+t^2+2u\,\mU}{st}\,,
    \label{eq:photon+qQED}
\end{equation}
where $Q_\ell$ and $Q_{\rm q}$ are lepton and quark electric charges, respectively. We note that the reported matrix elements should be divided by $\alpha_s/2\pi$ to adapt to the \texttt{POWHEG} input~\cite{Alioli:2010xd}. Additionally, in order to study a more general flavor structure,  Eqs.~\eqref{eq:g+lQCD}$-$\eqref{eq:photon+qQED} should be multiplied by $(|\beta_L^{i\alpha}|^2+|\beta_R^{i\alpha}|^2)/2$, and the sum over all quark and lepton flavor combinations participating in the production should be taken.

Furthermore, we have used the derived averaged matrix elements to perform the analytic computation of the partonic cross-section for the $U_1$ production. The computation resembles the one for the case of the scalar leptoquark in Ref.~\cite{Greljo:2020tgv} and we here give a way to translate those results to the case of the $U_1$. In addition to the universal replacement $y_{q\ell}\to g_4$, 
we find for the gluon-lepton collision the following partonic cross-section
\begin{equation}
     \hat{\sigma}_{g\ell}^{\text{vec}} = 2\hat{\sigma}_{g\ell}^{\text{sca}}+ \frac{\pi z g_4^2}{2 m_U^2}\frac{\alpha_s}{2\pi}T_R \frac{2(1-z)}{z}\,,
     \label{eq:gl_partonic}
\end{equation}
where $\hat{\sigma}_{g\ell}^{\text{sca}}$ can be found in Eq.\,(2.4) in~\cite{Greljo:2020tgv}. For the cross-section corrected by the virtual contributions and the real-gluon emission, we obtain
\begin{align}
    \hat{\sigma}_{q\ell}^{\text{vec}} = 2\hat{\sigma}_{q\ell}^{\text{sca}} - \frac{\pi z g_4^2}{2 m_U^2}\frac{\alpha_s}{2\pi}C_F \bigg[\frac{2}{3}(1-z)\nonumber\\
    +\left(\frac{2\pi^2}{3}-\frac{4\pi}{\sqrt{3}}+\frac{37}{12}-\frac{3}{2}L_R\right)\delta(1-z)\bigg]\,,
    \label{eq:ql_partonic}
\end{align}
where $\hat{\sigma}_{q\ell}^{\text{sca}}$ corresponds to Eq.\,(2.3) in~\cite{Greljo:2020tgv}. Finally, the partonic cross-section for the photon-initiated process reads
\begin{align}
     \hat{\sigma}_{\gamma\ell}^{\text{vec}} = 2\hat{\sigma}_{\gamma\ell}^{\text{sca}}&+ \frac{\pi z g_4^2}{2 m_U^2}\frac{\alpha_{\text{QED}}}{2\pi}\bigg[4 Q_{\text{q}}Q_{\text{LQ}}\log(z)\nonumber\\
     &+(Q_{\text{LQ}}^2+Q_{\text{q}}^2z)\frac{2(1-z)}{z}\bigg]\,, 
     \label{eq:gammaq_partonic}
\end{align}
where $Q_{\text{LQ}}$ is the $U_1$ electric charge. In Eqs.~\eqref{eq:gl_partonic}-\eqref{eq:gammaq_partonic}, the variable $z=m_U^2/\hat{s}$ should be integrated over to obtain the hadronic cross-section
\begin{equation}
\sigma(s) = 2\sum_{ij}  \int_{\xi}^{1}dy \, f_i(y)\int_{\xi/y}^1 dz\, \frac{\xi}{y z^2} f_j\left( \frac{\xi}{yz}\right)\hat{\sigma}_{ij}^{\text{vec}}\,.
\label{eq:hadronic_xsec}
\end{equation}
The functions $f_i$ and $f_j$ correspond to the PDFs of $i$ and $j$ partons, $\xi = m^2_{U}/s$ with $\sqrt{s}$ being the collider center of mass energy, and $y$ is the fraction of the proton momentum carried by the parton labeled by $i$. As before, to allow for different $U_1$ couplings, the expressions for $\hat{\sigma}^{\text{vec}}$ should be multiplied by $(|\beta_L^{i\alpha}|^2+|\beta_R^{i\alpha}|^2)/2$.

Let us also note that the impact of corrections due to new heavy dynamics proportional to $\alpha_4 = g_4^2/16\pi^2$ has been evaluated in~\cite{Fuentes-Martin:2019ign}. Here, we assume that both parameters, $m_U$ and $g_4$, are defined in the on-shell scheme with respect to the $\mathcal{O}(\alpha_4)$ corrections, such that the $U_1$ on-shell production does not receive any modification by definition. On the other hand, the non-trivial interplay between the low- and high-energy observables implies that $U_1$ contributions at low-energy will be modified by $\mathcal{O}(\alpha_4)$ corrections. For example, the preferred region for addressing the B-meson anomalies in the $g_4-m_U$ plane will be modified in the presence of new heavy dynamics. We take this into account in our phenomenological analysis in Sec.~\ref{sec:pheno}. 

Finally, we allow for the treatment of the finite-width effects which are especially important in the large $g_4$ regime. In particular, we assume that the only decay channels of $U_1$ are to massless fermions, such that its decay width reads
 \begin{equation}\label{eq:U1width}
     \Gamma = \frac{g_4^2}{48 \pi}\,m_U  \sum_{i, \alpha} \Big(2 \, |\beta_L^{i \alpha}|^2 + |\beta_R^{i \alpha}|^2\Big)\,.
 \end{equation}
We implement the effects of $\Gamma$ during the generation of signal event samples by using the Breit-Wigner  (BW) prescription for the $U_1$ propagator~\cite{Gigg:2008yc}. It approximates the $2\to 2$ matrix element by adding an extra integration over the invariant mass which is spread around $m_U$ according to the Breit-Wigner  distribution~\cite{Buonocore:2022msy}. In the code, this is achieved by setting the flag $\texttt{BWgen}$ to 1 in the $\texttt{POWHEG}$ input card. On the other hand, setting $\texttt{BWgen}$ to 0 gives the narrow-width approximation (NWA) for the leptoquark resonance. 
As anticipated, for larger values of $g_4$, the finite-width effects become important, and $\sigma_{\rm BW}/\sigma_{\rm NWA} - 1 \simeq \mathcal{O} (1)$ already for $m_U \gtrsim 2.5 \, \mathrm{TeV}$ and $g_4 \sim 1.5$. Moreover, since the PDFs are convoluted with the BW weight inside the phase-space integral, larger widths allow for probing the lower Bjorken-$x$ region of the PDFs. This region of the phase space compensates for the possible BW suppression, ultimately leading to a larger cross-section in comparison to the NWA estimate. Such effects regarding the breakdown of the NWA have already been pointed out in~\cite{Berdine:2007uv}.

\section{Phenomenology}\label{sec:pheno}

\subsection{NLO inclusive cross section}\label{subsec:incxsec}
\begin{figure}
    \centering
    \includegraphics[width=0.95\columnwidth]{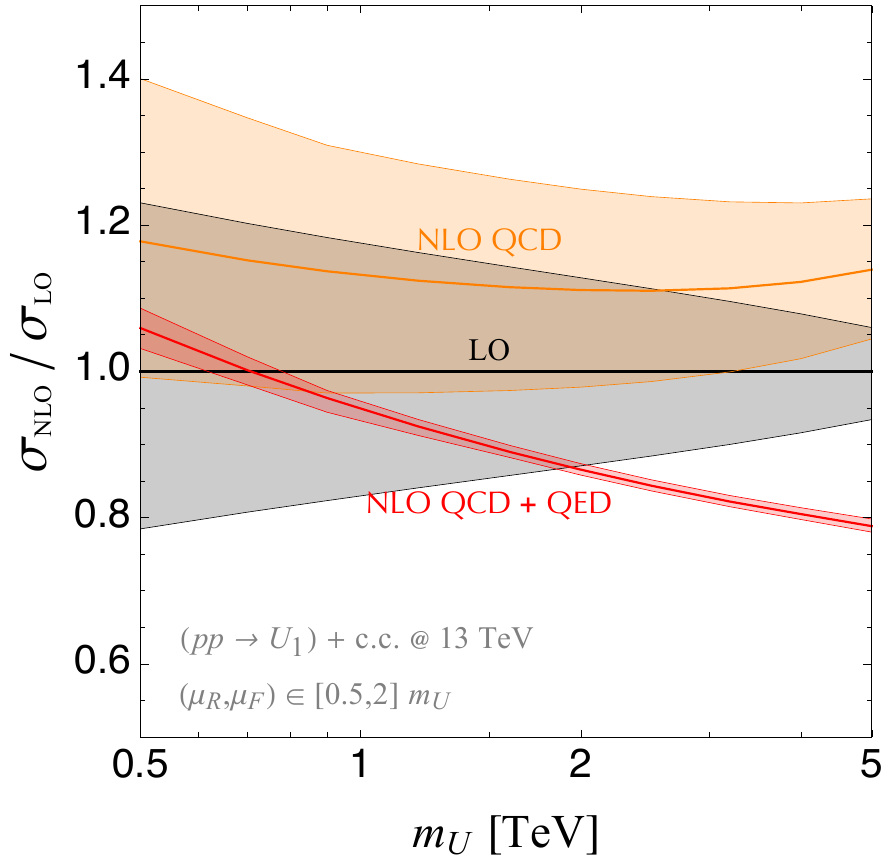}
    \caption{NLO $K$-factors for the resonant vector leptoquarks production at $\sqrt{s}=13\,\mathrm{TeV}$. The uncertainty bands were estimated using a seven-point scale variation.}
    \label{fig:kfactor}
\end{figure}

As a first phenomenological study, we compute the inclusive hadronic cross-sections at NLO QCD + QED for the resonant vector leptoquark production in $b+\tau$ collisions at the LHC. We set the energy of the beams to $6.5\,\mathrm{TeV}$ each and use the \texttt{LUXlep-NNPDF31\_nlo\_as\_0118\_luxqed} PDF set, which has been obtained by combining the \texttt{NNPDF} 3.1 PDF set \cite{Bertone:2017bme} together with the lepton PDFs reported in \cite{Buonocore:2020nai}. 
We scan over the relevant $U_1$ mass window, $m_U= [500,5000]$ GeV, and the coupling is set to $g_4 =1$. The entries of the $\beta_R$ and $\beta_L$ matrices are all set to zero with the exception of $\beta_L^{33}=\beta_R^{33} =1$, and for practical purposes we abbreviate $\beta_{L,R}^{33} \equiv \beta_{L,R}$. 
The results reported in Tab.~\ref{tab:x_secU1} are obtained in two ways. First, we integrate Eqs.~\eqref{eq:gl_partonic}-\eqref{eq:gammaq_partonic} over the appropriate PDFs, as in Eq.~\eqref{eq:hadronic_xsec}. Second, we let \texttt{POWHEG} with the flag \texttt{BWgen 0} perform the phase-space integrals with the input provided in Sec.~\ref{sec:resProd}. The two methods agree perfectly, which serves as a cross-check of our results.
To determine the scale uncertainties, we apply the typical seven-point scale variation, i.e. we set $\mu =\mu_R =\mu_F$ and then vary the two scales ($\mu_R$ the renormalization scale, and $\mu_F$ the factorization scale) by a factor of two. The uncertainty from the PDF set is estimated using the Monte Carlo replicas \cite{NNPDF:2017mvq, Bertone:2017bme}. Replicas are PDFs fitted to pseudodata, which is generated randomly using a Gaussian distribution around each data point with the experimental uncertainty as variance. With the reweighting procedure in \texttt{POWHEG-BOX}, both the scale variations and the reweighting to the replicas, can be achieved for each generated event.
 \begin{table}[t]
\renewcommand{\arraystretch}{1.4}
 \centering
 \begin{tabular}{|c|c|}
 \hline
 $m_U$\,[TeV] & $\sigma_{NLO}$\,[pb]\\
 \hline
 \hline
$0.50 $ & $ (2.67\cdot 10^{-1}) ^{+1.41 \% }  _{- 3.57 \% } \pm 2.00 \% $\\\hline
$1.00 $ & $ (1.14\cdot 10^{-2}) ^{+1.11 \% }  _{ -1.52 \% } \pm 2.02 \% $\\\hline
$1.50 $ & $ (1.29\cdot 10^{-3}) ^{+1.06 \% }  _{ -0.87 \% } \pm 2.38 \% $\\\hline
$2.00 $ & $ (2.20\cdot 10^{-4}) ^{+1.05 \% }  _{ -0.73 \% } \pm 2.93 \% $\\\hline
$2.50 $ & $ (4.69\cdot 10^{-5}) ^{+1.09 \% }  _{ -0.76 \% } \pm 3.67 \% $\\\hline
$3.00 $ & $ (1.14\cdot 10^{-5}) ^{+1.15 \% }  _{ -0.81 \% } \pm 4.75 \% $\\\hline
$3.50 $ & $ (3.01\cdot 10^{-6}) ^{+1.39 \% }  _{ -0.88 \% } \pm 6.19 \% $\\\hline
$4.00 $ & $ (8.27\cdot 10^{-7}) ^{+1.82 \% }  _{ -0.96 \% } \pm 8.09 \% $\\\hline
$4.50 $ & $ (2.32\cdot 10^{-7}) ^{+2.07 \% }  _{ -1.04 \% } \pm 10.89 \% $\\\hline
\end{tabular}
\caption{Inclusive cross-section for the $U_1$ leptoquark produced by $b + \tau$ collisions at NLO QED + QCD. The first (second) uncertainty is from seven-point scale variations (PDF replicas).}
 \label{tab:x_secU1}
\end{table}
\\\indent The NLO $K$-factors, defined as the ratio of the hadronic cross section after and before the NLO corrections, are plotted in Fig.~\ref{fig:kfactor}. The central solid lines are obtained for $\mu_R=\mu_F=m_U$, while the bands reflect the scale uncertainties from the seven-point variation. The QCD corrections are positive and are mostly independent of the $U_1$ mass. On the other hand, the QED corrections are negative and larger than the QCD ones for $m_U\gtrsim 700$ GeV. What stands out is the cancellation of scale uncertainties when including both QCD and QED corrections, emphasizing the effect of the photon-initiated process in reducing the theoretical uncertainties. As can be seen from Tab.~\ref{tab:x_secU1}, the leading source of theoretical uncertainties is inadequate knowledge of the PDFs for large $U_1$ masses. 

\subsection{Differential distributions}\label{subsec:diffdistro}

In this subsection, we study differential distributions obtained from our simulation for the $b + \tau \rightarrow U_1$ leptoquark production. 
The most natural choice for the discriminating variable in resonant production searches is the invariant mass of the final states. In our case, the presence of the $U_1$ can be inferred through the invariant mass of the final jet-lepton system. We study the result for leading order and next-to-leading order events both before and after the parton shower. The mass is set to $m_U = 2000 \,\mathrm{GeV}$ and we use the same couplings as in Sec.~\ref{subsec:incxsec}.  
Therefore, the possible decay products of the $U_1$ leptoquark are $b+\tau$ or $t+\nu_\tau$. 
The events are passed to \texttt{Herwig} \cite{Bellm:2019zci} for the parton shower and the anti-$k_T$ algorithm from \texttt{FastJet} \cite{Cacciari:2005hq, Cacciari:2011ma} is used for jet clustering. To enable the treatment of vector leptoquarks in \texttt{Herwig}, a suitable \texttt{FeynRules} \cite{Alloul:2013bka} UFO-model needs to be used, e.g. \cite{Dorsner:2018ynv, Baker:2019sli, Bhaskar:2021pml, Bhaskar:2020gkk, Bhaskar:2021gsy}.

In the benchmarks for the scalar leptoquarks reported in \cite{Buonocore:2022msy}, the reconstruction was done by picking the hardest jet and lepton. This will still work before the parton shower, but not after, since the $\tau$-lepton decays and produces a jet. A simple $\tau$-tagging procedure is used: each jet close to a $\tau$-lepton, with separation $\Delta R < 0.5$, is tagged with a static efficiency of $60\%$. However, just selecting the hardest $\tau$-tagged jet and the hardest $b$-jet is not enough to reconstruct the leptoquarks mass adequately, since there is a significant amount of missing transverse momentum due to neutrinos produced in the $\tau$-lepton decay. Therefore, we implement a simple recast analysis that relies on the cuts given in the mentioned CMS leptoquark search \cite{CMS:2023bdh}. 
These cuts are selected to reduce most of the SM background events, characterized by softer radiation. 

For the transverse momentum of the selected lepton or $\tau$-tagged jet we set $p_T ^\ell > 200\,\mathrm{GeV}$, and $p_T ^j > 300\,\mathrm{GeV}$ for the jet associated with the quark.
A cut of $p_T ^{miss} > 100\,\mathrm{GeV}$ is set for the missing transverse momentum, and for the transverse momentum $p_T ^{\ell + miss} > 100 \,\mathrm{GeV}$ of the $p_T ^{miss} + p^\ell $ system. Rapidity cuts are set to $\eta _j < 2.4$ and $\eta _\ell <2.1$. 
To capture the missing transverse momentum of the neutrinos produced during the $\tau$ decay, the azimuthal angle constraint between the visible part of the lepton and the missing $p_T$, $\Delta \phi _{\ell,miss} < 0.3$, is set.  Moreover, the jets selected as the decay products of the leptoquark should be separated by demanding a cut on $\Delta R > 0.5$.

\begin{figure}
    \centering
    \includegraphics[width=\columnwidth]{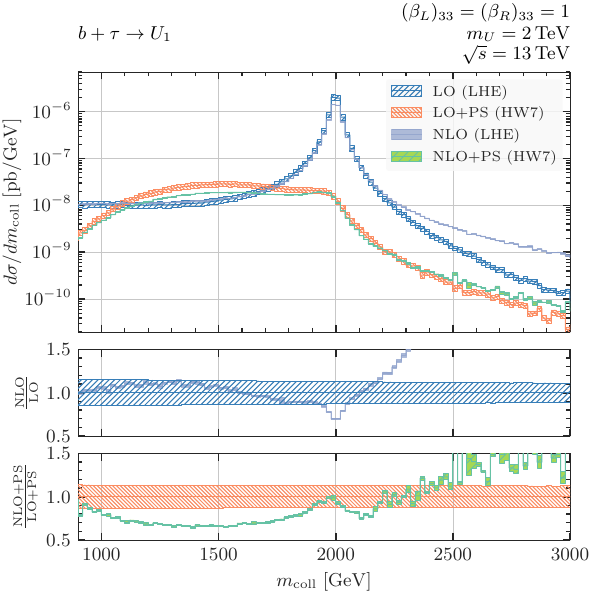}
    \caption{Differential distribution for $m_{\textrm{coll}}$ for the benchmark process described in Sec.~\ref{sec:pheno} in the upper panel. The ratio NLO to LO of unshowered (showered) events is in the second (third) panel. The error bands were obtained using a seven-point scale variation. A $\tau$-lepton is considered stable at the level of the LHE files, while its decay is implemented in the PS.}
    \label{fig:mcoll}
\end{figure}

To reconstruct the mass of the leptoquark, we use the collinear mass defined as~\cite{CMS:2023bdh}
\begin{equation}
    m_{\text{coll}} = \displaystyle m_{\text{vis}} \sqrt{\frac{p_T ^{\text{vis}} + p_T ^{\text{invis}}}{p_T^{\text{vis}}}}\,,
    \label{eq:m_coll}
\end{equation} where $p_T ^{\mathrm{invis}}$ is the part of $p_T ^{\mathrm{miss}}$ in direction of the $\tau$ decay products. 
Since the decay of the $\tau$-lepton is handled by \texttt{Herwig}, the quantity $p_T ^{\mathrm{invis}}$ will only be non-zero for showered events. 
Therefore, if events are analyzed before the parton shower, the collinear mass corresponds to the invariant mass of the $b + \tau$ pair. 
All cuts mentioned above, as well as the couplings, can be modified in the \texttt{POWHEG} input card. 
\\\indent The resulting differential distribution, $d\sigma/dm_{\text{coll}}$, is shown in Fig.~\ref{fig:mcoll}. A substantial difference is observed when the parton shower is turned on. In particular, after the parton shower, the $\tau$-tagging and the additional cuts mentioned above cause fewer events to be reconstructed. Nevertheless, the differential cross-section is found to be larger after the shower for $m_{\text{coll}}$ values below the peak. This is due to additional final-state radiation that lies outside the $b$- and $\tau$-jet cones. The same effect also lowers the other tail of the distribution. Moreover, in the case of showered events, more jets are present, and the chance of misidentifying is higher. 
The NLO distribution, without the parton shower, has a broadened profile as a consequence of additional radiation in comparison to the leading order case.
This slight depletion of the resonance peak can best be seen in the ratio subplot in Fig.~\ref{fig:mcoll}.

\begin{figure}[t]
    \centering
    \includegraphics[width=\columnwidth]{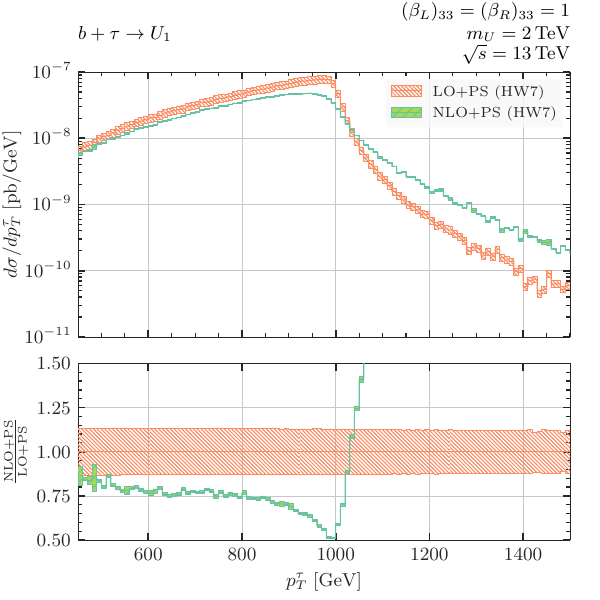}
    \caption{Differential distribution for the transverse momentum of the tau-tagged jet with missing $p_T$ and the ratio NLO to LO below. This quantity is only computed after the parton shower since the $\tau$-lepton does not decay otherwise. The error bands were computed using the seven-point prescription for scale variation.} 
    \label{fig:pttaujet}
\end{figure}

As a different application of the code, in Fig.~\ref{fig:pttaujet}, we show the differential distribution of the transverse momentum for the tau-tagged jet. The missing $p_T$ in the direction of the tau-tagged jet was added to obtain the expected Jacobian peak at $m_{U}/2= 1000\,\mathrm{GeV}$. Since finite-width effects were taken into account in our simulation, the peak is slightly smeared out. In the next-to-leading-order case, the extra radiation causes the leptoquark to have a non-zero transverse momentum, which causes the tail of the distribution above $m_U /2$ to be raised compared to the leading-order case. 
\\\indent We note that differential distributions for other kinematic quantities can easily be obtained by running the code. 

\subsection{LHC bounds}
\label{subsec:Projected_bounds}

Given the available CMS search results \cite{CMS:2023bdh} which, among others, target $b + \tau$ final states emerging from a scalar leptoquark $\mathrm{LQ}_s$, we can establish constraints on the coupling strength $g_4$ in relation to the leptoquark mass $m_U$.
The search relies on proton-proton collision data at a center-of-mass energy of $\sqrt{s} = 13$  TeV, obtained from the CMS detector, corresponding to an integrated luminosity of $\mathcal{L} = 138 \, \mathrm{fb}^{-1}$.
To this end, it is necessary to highlight any potential differences that must be considered while applying this analysis to the $U_1$ signal. 
In general, the efficiency, i.e. the signal acceptance rate of the $U_1$ production for the cuts presented in Sec.~\ref{subsec:diffdistro} can be different. By performing a cut-and-count analysis for both scalar and vector leptoquark cases, we find that the efficiencies exhibit a minimal difference. This outcome is expected, given that the experimental search focuses on reconstructing signal events from the resonance peak, and other kinematical details become irrelevant. 
\\\indent The exclusion bounds, for different values of the scalar leptoquark coupling $\lambda$, are available on \texttt{HEPData}~\cite{hepdata.141335}, where the coupling $\lambda$ and the partonic cross-section for the scalar leptoquark production at LO are related as~\cite{Buonocore:2022msy}
\begin{equation}\label{eq:PHENO:sLQxsec}
\hat{\sigma}_{\mathrm{LQ_s}} = \frac{\lambda^2}{4} m_{\rm LQ_s}^2 \,.
\end{equation}
The experimental sensitivity decreases for larger values of $\lambda$ due to the increase in the leptoquark width, while the experimental resolution of $m_{\rm coll}$ is fixed. Therefore, we can directly translate the exclusion limits from the scalar to the vector leptoquark case, provided the widths are the same for both choices of couplings. 
This is achieved once the couplings ($g_4, \beta_L, \beta_R$) satisfy the following relation for a fixed $\lambda$
\begin{equation}\label{eq:PHENO:lambdag4width}
\lambda=\frac{g_4}{\sqrt{3}}\sqrt{2 |\beta_L|^2 + |\beta_R|^2}\,.
\end{equation}

\begin{figure*}[t!]
    \centering
    \includegraphics[width = 1.03\columnwidth]{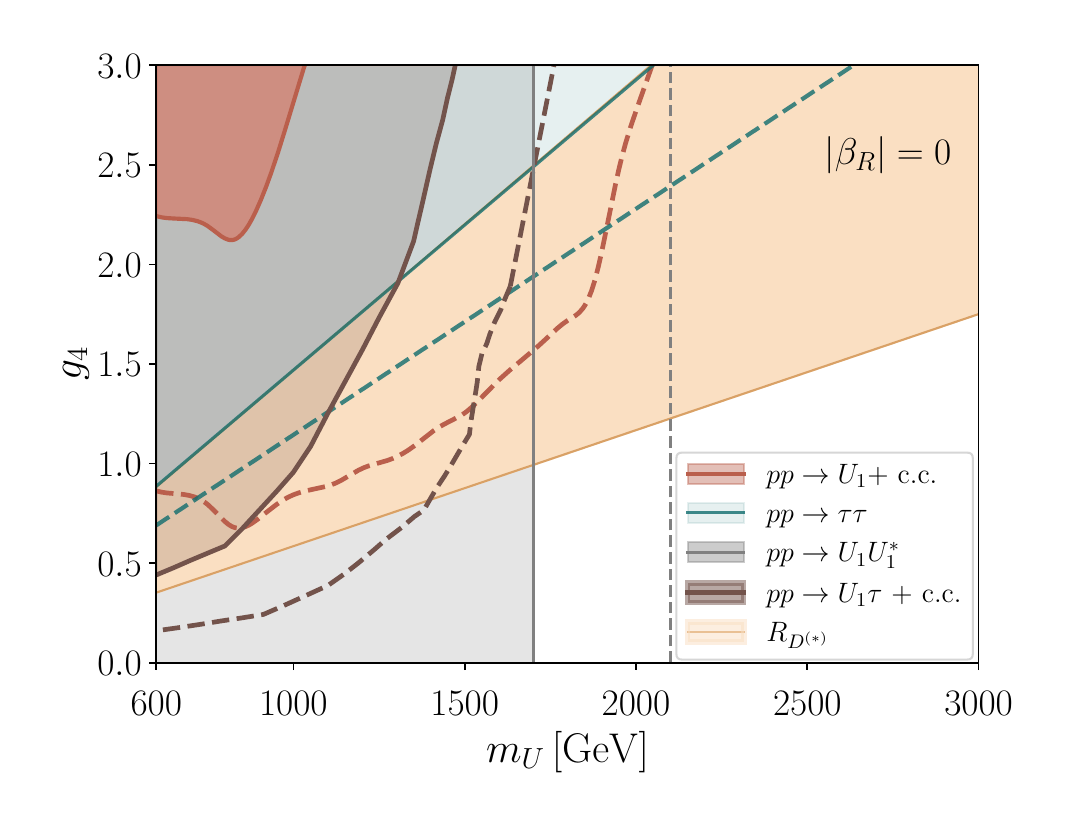}
    \includegraphics[width =1.03\columnwidth]{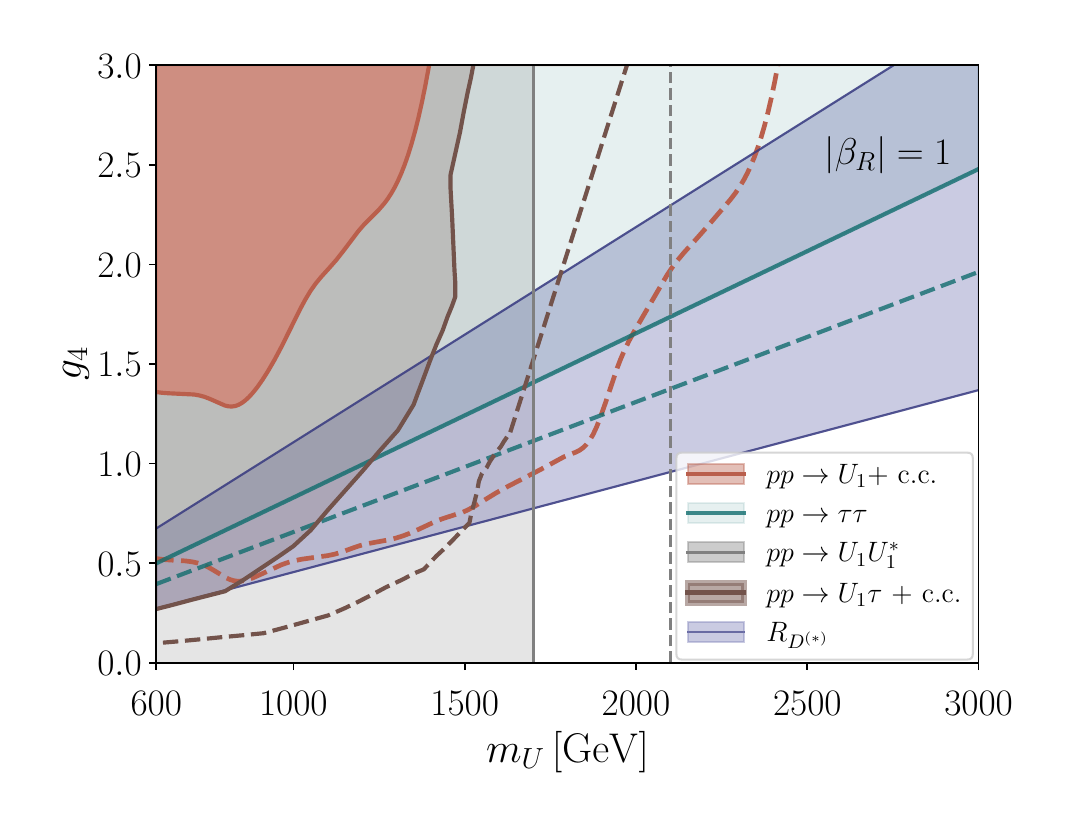}
    \caption{Excluded parameter space at $95\%$ CL for the $U_1$ resonant production from the observed limits, shown as the red region, in comparison to other production mechanisms. The bounds from $pp \to \tau \tau$ 
    at 137 $\mathrm{fb}^{-1}$ derived by CMS \cite{CMS:2020wzx}, are depicted by the solid green line. The brown area is the excluded parameter space from the single leptoquark production~\cite{CMS:2023qdw}. The gray line corresponds to the exclusion limits set by the leptoquark pair production $pp \to U_1 U_1^\dagger$ \cite{CMS:2020wzx}. The dashed lines represent the expected bounds, projected to 3~$\mathrm{ab}^{-1}$ for the c.o.m. energy $\sqrt{s} = 14$ TeV and follow the same color coding. In the left (right) panel, $|\beta_R|$ is set to zero (one), whereas $|\beta_L|$ equals 1 in both cases.}
    \label{fig:PHENO:bounds}
\end{figure*}We employ \texttt{POWHEG} to compute the $U_1$ production cross section at NLO, using the full BW prescription, and assume that the leptoquark solely couples to third-family fermions. We analyze two distinct cases, namely $|\beta_L| = |\beta_R| = 1$ and $|\beta_L| = 1, |\beta_R| = 0$, resulting in a fixed branching ratio of $68 \% $ ($ 50 \% $) for the first (second) choice. We note that in the latter case, the $U_1$ resonance is narrower compared to the scalar leptoquark case if we take $\lambda = g_4$. The slight improvement in sensitivity, however, is found to be negligible. 
\\\indent Calculating the cross-section for various points in the $(g_4, m_U)$ parameter space, we can determine the values that saturate the imposed limits. In Fig.~\ref{fig:PHENO:bounds}, we compare the bounds resulting from the resonant production of $U_1$ in $b+\tau$ fusion, and the bounds derived from alternative vector leptoquark production mechanisms for the choices $|\beta_{L}| = |\beta_{R}| = 1$ and $|\beta_L| = 1, \, |\beta_R| = 0$. 
The current exclusion bounds from lepton-quark fusion are obtained from translating the observed CMS bounds for the scalar leptoquarks as discussed above, and are shown in the red region in Fig.~\ref{fig:PHENO:bounds}. 
The most stringent exclusions arise from the non-resonant effect of $U_1$ in $pp \to \tau \tau$~\cite{ATLAS:2020zms,CMS:2022rbd,Aebischer:2022oqe} shown in the green area. In the region of small leptoquark masses, the dominant bounds arise from the pair production~\cite{CMS:2020wzx} given by the gray area.
Additionally, we display the bounds from $pp \to U_1 \tau$, where the excluded portion of the parameter space is represented by the brown region. 
The dashed lines represent projections of the expected bounds at the high-luminosity LHC phase (HL-LHC) reaching $\mathcal{L} = 3 \, \mathrm{ab}^{-1}$ with the center-of-mass energy  $\sqrt{s} = 14$ TeV~\cite{ZurbanoFernandez:2020cco}. For the derivation of these bounds, we consider that the experimental uncertainty is dominated by statistics, hence the sensitivity scales with the square root of the luminosity. The projected limits for the single production channel are extrapolated assuming the same sensitivity for masses above $m_U > 1600$ GeV as for $m_U = 1600$ GeV, given in \cite{CMS:2023qdw}. The lepton-quark fusion projections improve better relative to the single production ones owing to the larger discrepancy between observed and expected limits in the former production channel~\cite{CMS:2023bdh,CMS:2023qdw}.\\
\indent Apart from the exclusion bounds comparison, we also check the sensitivity of the resonant channel to the 90\% CL preferred region addressing the hints of LFU violation in the observables $R_D$ and $R_{D^\ast}$~\cite{Aebischer:2022oqe}. 
As can be seen, the lepton-quark fusion will help to scrutinize the low-energy parameter space preferred by the charged-current B-meson anomalies during the HL-LHC.\\
\indent Although the present $b+\tau$ fusion search does not exclude any additional part of the $g_4 - m_U$ parameter plane, the exclusion limits are competitive to other production mechanisms at the HL-LHC. Specifically, for $g_4 \gtrsim 1$, the resonant production channel outperforms the single production mechanism. When comparing to the non-resonant $pp \to \tau \tau$ process, it is important to note that the sensitivities scale differently with the luminosity, $\mathcal{L}$, improvement. 
For a fixed branching ratio, the resonant production cross-section scales with the coupling squared, which means that the bounds on the coupling scale as $\mathcal{L}^{-1/4}$. In contrast, the bounds from $pp \to \tau \tau$ will improve by an approximate factor of $\mathcal{L}^{-1/8}$. Therefore, we emphasize the important role that the lepton-quark fusion will play during the HL-LHC: it provides complementary information about the $U_1$ leptoquark that experimental collaborations at the LHC should explore.
\indent Finally, we note that phenomenologically allowed departures from the minimal model discussed here have a minuscule impact on the derived bounds. As an example, it is possible to achieve $U_1$ couplings to light-family fermions through mixing with additional fermion states which are vector-like under the SM gauge group~\cite{Fuentes-Martin:2020hvc}. In particular, a fit to low-energy observables in~\cite{Cornella:2021sby} prefers $\beta_L^{23}\simeq 0.2$, which results in an additional production channel for the $U_1$ leptoquark. We have checked that including such contribution from $s+\tau$ collisions results in an improvement of $\simeq 2.5\%$ in the exclusion bounds. In the case of other couplings, the situation is worse as the PDF enhancement for lighter quarks cannot compensate for the additional coupling suppression as dictated by the model's $U(2)^5$ protection and strong constraints on the light-family $U_1$ couplings. Thus, we expect the reported bounds to be robust to model modifications and represent a large class of models based on the $\mathcal{G}_{4321}$ gauge group.

\section{Conclusions}
The main objective of this work is to give a precise analysis of a direct $b+\tau$ fusion producing a vector leptoquark at the LHC. Such a state is part of the spectrum of NP constructions that feature third-family quark-lepton unification. Being interested in the NLO analysis, we work with a full UV model based on a flavor non-universal gauge group $\mathcal{G}_{4321}$ with an $SU(4)$ factor acting on the third-family fermions as in Tab.~\ref{tab:fields}. As emphasized in the introduction, this class of models has been identified as a possible last step in breaking to the SM, of more ambitious models addressing the $U(2)^5$-like structure of the SM Yukawas, the electroweak scale stability, and the charge quantization problem. 

After describing the gauge boson dynamics, we presented the results of NLO QCD + QED corrections for the partonic processes contained in $pp \to U_1$. We found that two corrections partially cancel, resulting in a smaller cross-section for a wide range of vector leptoquark masses, as shown in Fig.~\ref{fig:kfactor}. Our computation led to a substantial reduction of the renormalization and factorization scale variation uncertainties, with the leading source of theoretical error being the limited knowledge
of PDFs at large $x$, indicated in Tab.~\ref{tab:x_secU1}.\\
\indent Following the implementation of the NLO results into \texttt{POWHEG}, together with the utilization of the parton + lepton showering algorithm offered by \texttt{Herwig}, we have effectively constructed a Monte Carlo event generator for this production channel. It can be obtained from the following Github repository \href{https://github.com/peterkrack/3rd-Lepton-Quark-Fusion}{\texttt{github.com/peterkrack/3rd-Lepton-Quark-Fusion}}, and used to perform detailed phenomenological analyses by studying arbitrary differential distributions. As an example, we produced the distribution of events in a discriminating observable $m_{\text{coll}}$ (see Eq.~\eqref{eq:m_coll}) used in the CMS search for scalar leptoquarks in this channel~\cite{CMS:2023bdh}.
Based on our findings, we were able to translate the CMS exclusion bounds for the scalar leptoquarks to the case of the $U_1$.
\\\indent Despite the limiting exclusion power of the lepton-quark fusion channel at present, we have demonstrated that this channel will be prominent during the high-luminosity LHC phase. Owing to the resonant enhancement, the limits on the $U_1$ leptoquark parameter space from the lepton-quark fusion are more sensitive to luminosity and the center-of-mass energy improvements than the limits from the other channels, resulting in their full complementarity in Fig.~\ref{fig:PHENO:bounds}. Our work provides the necessary ingredients on the theoretical side to make the best use of this, opening the possibility of performing the first experimental search for the $U_1$ vector leptoquark in the lepton-quark fusion channel. \\

\section{Acknowledgements}
We thank Javier M. Lizana and Jernej F. Kamenik for useful discussions, and Admir Greljo for carefully reading the manuscript. This project has received funding from the INFN Iniziative Specifica APINE, the Slovenian Research Agency (grant No. J1-3013 and research core funding No. P1-0035), and the European Research Council (ERC) under the European Union's Horizon 2020 research and innovation programme under grant agreement 833280 (FLAY).

\appendix

\bibliographystyle{ieeetr}
\bibliography{references.bib}

\end{document}